\documentclass[aps, prl,longbibliography, 
amsmath,amssymb,
 reprint,%
]{revtex4-2}

\usepackage{graphicx}
\usepackage{dcolumn}
\usepackage{dsfont}
\usepackage{bm}

\usepackage[T1]{fontenc}
\usepackage{etoolbox}
\usepackage{xspace}
\usepackage{array}
\usepackage{siunitx}
\usepackage{booktabs}
\usepackage{bigstrut}
\usepackage{stmaryrd}

\usepackage{longtable} 
\usepackage{tabularx} 

\newcolumntype{K}[1]{>{\centering\let\newline\\\arraybackslash\hspace{0pt}}m{#1}}
\usepackage{makecell}
\usepackage{float}

\usepackage[hidelinks]{hyperref}
\hypersetup{
  colorlinks   = true, 
  urlcolor     = blue, 
  linkcolor    = blue, 
  citecolor   = blue 
}
\usepackage{soul}

\makeatletter
\def\@email#1#2{%
 \endgroup
 \patchcmd{\titleblock@produce}
  {\frontmatter@RRAPformat}
  {\frontmatter@RRAPformat{\produce@RRAP{*#1\href{mailto:#2}{#2}}}\frontmatter@RRAPformat}
  {}{}
}%

\usepackage[mathscr]{euscript}
\usepackage{color}
\usepackage{xcolor}
\usepackage{amsmath}
\usepackage[bb=boondox]{mathalfa}
\usepackage{mathrsfs}
\usepackage{multirow}

\usepackage{mathtools}

\DeclarePairedDelimiter\floor{\lfloor}{\rfloor}

\makeatother
\begin{document}

\preprint{APS/123-QED}

\title{Cyclicity of interaction frame transformations}

\author{Michael C. D. Tayler*}
\affiliation{ICFO—Institut de Ciències Fotòniques, The Barcelona Institute of Science and Technology, Castelldefels
(Barcelona) 08860, Spain.}
\email{michael.tayler@icfo.eu}

\author{Mohamed Sabba}
\affiliation{School of Chemistry, University of Southampton, SO17 1BJ, United Kingdom.}

\date{\today}

\begin{abstract}
We identify a cyclic property of rotation sequences involving piecewise displacements $\beta$ about arbitrary axes in three dimensions.  Specifically, when transformation to the toggling frame is applied successively $m$ times, for $\beta=2\pi/m$ the original sequence returns.  This main result unifies several families of rotation sequences used for error-tuned control across quantum technologies, from NMR and MRI to atomic clocks and atom-scale computing.  For the widest class of cycle, $m=2$, we highlight sequence \textit{duality} where every narrowband $\pi$-element sequence has a broadband $\pi$-element counterpart, and vice-versa.  Higher cycles ($m>2$) connect to polyhedral models for error-tolerant sequence design, characterized by vertex axes with $m$-fold rotational symmetry.  We derive original sequences and outline their applications to spin control and spin decoupling.
\end{abstract}
                              
\maketitle

\textit{Introduction}---Rotational periodicity arises in a wide variety of systems whose configuration spaces are discrete rather than continuous. In one dimension, examples include gear trains, Möbius strips, and spinor behavior such as the double cover of SO(3) by SU(2). In higher dimensions, notable cases include twisty puzzles like the Rubik’s cube \cite{bao_twistypuzzleinspired_2024} and quantum systems exhibiting geometric phase \cite{skrynnikov_classification_1994}, where repeated sequences of noncommuting operations restore the system to its initial state. Recognizing the cyclic structure of these systems provides a perspective on their underlying symmetries and has implications for algorithm design, especially in quantum control \cite{cai_quantum_2023}.

In this letter, we introduce and examine a rotational periodicity property that is intrinsic to the so-called \textit{toggling frame} -- the interaction frame defined as co-rotating with a piecewise-time-dependent torque or Hamiltonian applied to a system -- that is bread-and-butter to dynamics analysis, qubit dynamical decoupling \cite{souza_robust_2012,genov_arbitrarily_2017,choi_robust_2020}, noise filtering \cite{merrill_transformed_2014,khodjasteh_performance_2007}, and optimal control \cite{khaneja_optimal_2005}.  A context where this periodicity can be applied is \textit{composite} rotation sequences with tuned resilience to control field imperfections and inhomogeneities  \cite{souza_robust_2011,wolfowicz_pulse_2016,vandersypen_nmr_2005,jones_quantum_2011,jones_controlling_2024}.  
Composite rotations are use widely for high-fidelity manipulation of spin qubits, for instance, in nuclear magnetic resonance (NMR) \cite{levitt_composite_1986,levitt_composite_2007,jones_controlling_2024}, where this field began, in molecular spectroscopy \cite{warren_multiple_1983,warren_there_1986}, and more recently in photonic quantum memories \cite{gupta_preserving_2015}, quantum computing \cite{cummins_tackling_2003,jones_quantum_2011} in excited nuclear states \cite{amiri_composite_2023}, trapped ions \cite{mallweger_motionalstate_2024,zanon-willette_composite_2016,zanon-willette_composite_2018,merrill_transformed_2014}, trapped neutral atoms \cite{lundblad_fieldsensitive_2009,dunning_composite_2014}, color centers \cite{aiello_compositepulse_2013}, and superconducting transmons \cite{collin_nmrlike_2004,torosov_experimental_2022,williams_quantification_2024,kuzmanovic_highfidelity_2024}. Analogous sequences may apply to classical systems with limited rotational-control degrees of freedom, such as spherical wheels and rolling robots \cite{diouf_spherical_2024,he_underactuated_2019}. 

\textit{Multiple toggling frames}---The main result is as follows: for a sequence of $n$ rotations through uniform angle steps, $\beta = 2\pi/m$, about arbitrary axes in three dimensions, denoted $\bm{e}_i^{(0)}$ ($i \in \{0:n-1\}$), the transformation to the toggling frame is periodic; we find that $m$ successive transformations returns to the original frame.  
From a physical perspective this result is remarkably simple, yet is unexpected and non-intuitive due to the inherent non-commutativity of 3D rotations.  However, it follows naturally from the cumulative definition of the frame.

Let us show this result for the sequence of arbitrary rotations, denoted as $\mathcal{S}^{(0)} \equiv \{ (\beta, \boldsymbol{e}_0^{(0)}), \ldots , (\beta, \boldsymbol{e}_{n-1}^{(0)})\}$, with left-to-right chronology.  
If a single rotation operator is described by $ R(\beta_i, \bm{e}_i^{(0)})$, then the overall rotation is described by the propagator
\begin{eqnarray}
    U_n{(\mathcal{S}^{(0)})} &=& \prod_{j=0}^{n-1} R(\beta, \bm{e}_j^{(0)}) \label{eq:Udef}\\
    &\equiv& R(\beta, \bm{e}_{(n-1)}^{(0)}) \ldots R(\beta, \bm{e}_1^{(0)}) R(\beta, \bm{e}_0^{(0)})\,, \nonumber
\end{eqnarray}  
which by closure of the rotation group is also a rotation.

As shown in Appendix A, erroneous rotations through $\beta\rightarrow\beta+\epsilon$ manifest in a toggling frame with dynamics given by rotations through angles $\epsilon$ about a set of axes $\bm{e}_i^{(1)}$ ($i \in \{0:n-1\}$), which are defined by the action of the inverse of $U_i(\mathcal{S}^{(0)})$ on each $\bm{e}_i^{(0)}$; note the subscript $i$, which implies cumulative dynamics:
\begin{eqnarray}
    \bm{e}_i^{(1)} &=& U_i^{-1}{(\mathcal{S}^{(0)})}\,\bm{e}_i^{(0)}; \qquad i\geq1\,,\label{eq:etilde} \\
    \bm{e}_0^{(1)} &=& \bm{e}_0^{(0)}\,. \nonumber
\end{eqnarray}
Coherent averaging theory states that the first-order effect of $\epsilon$ is zero if the mean (or centroid) 
\begin{equation}
\bm{C}^{(1)} = (1/n)\sum_{i=0}^{n-1}\bm{e}_i^{(1)} \label{eq:Cdefn}
\end{equation}
of $\{\bm{e}_i^{(1)}\}$ is zero; this is a target for broadband sequence design.  For more information, see Appendix B. 

From here, we will denote the toggling frame transformation as a map on $\{\bm{e}_i^{(0)}\}$ using the shorthand $\{\bm{e}_i^{(1)}\} = \hat{M} \{\bm{e}_i^{(0)}\}$, where $\hat{M}$ is defined by \autoref{eq:etilde}.  See also Appendix C.
When $\hat{M}$ is applied iteratively $m$ times, subsequent sets of vectors $\{\bm{e}_i^{(m)}\}$ are generated by concatenating rotations around axes $\{\bm{e}_j^{(m-1)}\}, (j\leq i)$:
\begin{equation}
    (\{\bm{e}_0^{(m)}, \bm{e}_1^{(m)}, \ldots, \bm{e}_{(n-1)}^{(m)}\}) = \hat{M}^m (\{\bm{e}_0^{(0)}, \bm{e}_1^{(0)}, \ldots, \bm{e}_{(n-1)}^{(0)}\})\,. \label{eq:Mpowerk}
\end{equation}
Remarkably, despite the folded nature of $\hat{M}$, each transformed vector $\bm{e}_i^{(m)}$ can also be expressed simply, in terms of the initial set:
\begin{equation}
    \bm{e}_i^{(m)} = \left(\prod_{j=0}^{i-1}{R(m\beta,\bm{e}_j^{(0)})} \right)^{-1}\bm{e}_i^{(0)}\,. \label{eq:Rpowerk}
\end{equation}
For example, $\bm{e}_3^{(m)} = 
U_3^{-1}{(\mathcal{S}^{(m-1)})}\ldots\,
U_3^{-1}{(\mathcal{S}^{(0)})}\bm{e}_3^{(0)}
=$ $R(-m\beta,\bm{e}_0^{(0)})R(-m\beta,\bm{e}_1^{(0)})$ $R(-m\beta,\bm{e}_2^{(0)}) \bm{e}_3^{(0)}$.

\autoref{eq:Rpowerk} now reveals the periodic property.  For rotation angles that are unit fractions of $2\pi$, ($\beta = 2\pi/m$), the net effect of $\hat{M}^m$ is an identity operation, returning the original vectors regardless of orientation: $\{\bm{e}_i^{(m)}\} \equiv \hat{M}^m \{\bm{e}_i^{(0)}\} = \{\bm{e}_i^{(0)}\}$.  In these cases, we say $\hat{M}$ has a cycle of length $m$.  The inverse transformation also becomes trivial, we see $\hat{M}^{-1}\equiv\hat{M}^{m-1}$ because $\hat{M}^{m-1}\hat{M}=1$. 

\textit{Application to composite pulses}---Despite wide application of the toggling frame to dynamical systems, \autoref{eq:Rpowerk} as a general property appears unrecognized.  A very limited use of the formula for $m=2$ (an involution) and equatorial-plane vectors $\{\bm{e}_i^{(0)}\}=\{(\cos \phi_i^{(0)},\sin \phi_i^{(0)},0)\}$ appears in the NMR literature \cite{wimperis_composite_1989,wimperis_iterative_1991,odedra_dualcompensated_2012}, in relation to the rf and toggling-frame phases of $\beta=\pi$-composite pulses. Phases of the rf pulses are related by the formulae 
\begin{eqnarray}
    \phi_i^{(1)} &=& \phi_0^{(0)} + \sum_{j=1}^{i}(-1)^j(\phi_j^{(0)}-\phi_{j-1}^{(0)})\,,\label{eq:phasetotildeframe}\\
    \phi_i^{(0)} &=& \phi_0^{(1)} + \sum_{j=1}^{i}(-1)^j(\phi_j^{(1)}-\phi_{j-1}^{(1)})\,,\label{eq:phasefromtildeframe}
\end{eqnarray}
where vectors $\{\bm{e}_i^{(1)}\}=\{(\cos \phi_i^{(1)},\sin \phi_i^{(1)},0)\}$ result directly from \autoref{eq:Rpowerk} because any exact $\pi$ rotation involving two rf-frame axes, $R(\pi,\bm{e}_i^{(0)})\,\bm{e}_k^{(0)}$, keeps the vector $\bm{e}_i^{(1)}$ confined to the $xy$ plane. 
The phase relation for $m=2$ was also mentioned by Pines \textit{et al.}\ in relation to the fixed-point theory of iterative dynamical decoupling \cite{tycko_iterative_1984,tycko_broadband_1984,tycko_fixed_1985}.  

Aside from these works, \autoref{eq:Rpowerk} is unexplored, especially for values $m>2$ and axes outside the equatorial plane.  An immediate conclusion is that there are $m$ sequences of rotations-by-$2\pi/m$ that cyclically interconvert with one another upon successive $\hat{M}$ transformations.  For $m=2$, this means a `duality' property of $\pi$ rotation sequences, where the toggling-frame vectors of one sequence equal the non-toggling (e.g., rf-frame) vectors of the dual sequence \footnote{We note, a constant phase offset is also allowed because $\hat{M}$ commutes with rotations about the $z$ axis.}.  As a general rule, a broadband compensated sequence $\mathcal{S}^{(0)}$ involving vectors $\{\bm{e}_i^{(0)}\}$ maps to a narrowband sequence $\mathcal{S}^{(1)}$, involving rotations about vectors $\{\bm{e}_i^{(1)}\}$, and vice versa. 
One interpretation is through the centroid of the vector sets.  Sequences for broadband angle-error compensation have balanced vectors $\bm{C}^{(1)}=\bm{0}$ and a far-from-zero $\bm{C}^{(0)}$, while for narrowband sequences it is the opposite: $\bm{C}^{(0)}=\bm{0}$ and a far-from-zero $\bm{C}^{(1)}$.  

The interchange of $\bm{e}_i^{(0)}$ and $\bm{e}_i^{(1)}$ between dual sequences also equates to a glide-reflection relationship in the profile of vector inversion vs.\ rotation angle.  If $q_{\mathcal{S}^{(0)}}(\bm{e}_\xi,{\beta'})$ is a function describing the transformation amplitude of a unit vector $\bm{e}_\xi$ under $\mathcal{S}^{(0)}$ with $\beta=\beta'$, where all rotation axes lie in the $xy$ plane, 
\begin{equation}
q_{\mathcal{S}^{(0)}}(\bm{e}_\xi,{\beta'})
= 
{\rm{Tr}} \{ ( \bm{e}_\xi U_{n}(\mathcal{S}^{(0)}))^{\dag} U_{n}(\mathcal{S}^{(0)}) \bm{e}_\xi \}\,,
\end{equation}
then provided that constraints $q_{\mathcal{S}^{(0)}}(\bm{e}_\xi,\,0) = 1$ and 
$q_{\mathcal{S}^{(0)}}(\bm{e}_\xi,\,\pi) = -1$ are satisfied (which describe nominal inversion), we find
\begin{equation}
q_{\mathcal{S}^{(1)}}(\bm{e}_\xi,{\beta'}) = -q_{\mathcal{S}^{(0)}}(\bm{e}_\xi,{\pi\pm\beta'})
\,. \label{eq:glidereflection}
\end{equation}

For $\bm{e}_\xi = \bm{e}_z \equiv (0,0,1)$, \autoref{eq:glidereflection} means the profiles of vector inversion vs.\ $\beta'$ for two dual sequences are related by a vertical flip and a translation by $\pi$ along the $\beta'$ axis, modulo $2\pi$.  The horizontal shift of $\pi$ is the period of $\beta$ for $m=2$, and the vertical flip, for one half of the dual, is supplied by the minus sign in $U_n(\mathcal{S}^{(k)})\bm{e}_z = -\bm{e}_z$ on the right-hand side of \autoref{eq:etilde}.  

Another general property of duals is parity reversal, where if the phase list $\{\phi_i^{(0)}\}$ is order-reversal antisymmetric then the list $\{\phi_i^{(1)}\}$ is order-reversal symmetric, and vice-versa.  The property arises from the alternating sign structure of \autoref{eq:phasetotildeframe}, which causes phase differences in the second half of the list to be the opposite of those in the first half.  This may be seen more clearly by stating \autoref{eq:phasetotildeframe} in a recursive form
\begin{equation}
\phi_{i}^{(1)} + (-1)^{i-1} \phi_{i}^{(0)} = \phi_{i-1}^{(1)} + (-1)^{i-1} \phi_{i-1}^{(0)}\,,
\end{equation}
or
\begin{equation}
\Delta_{-1} \phi_{i}^{(1)} = (-1)^{i} \Delta_{-1} \phi_{i}^{(0)}\,, \label{eqn:DualityOfDifferences}
\end{equation}
by introducing the finite difference operator $\Delta_a f(x) = f(x+a) - f(x)$.  

\begin{figure}[b]
    \centering
    \includegraphics[width=0.7\columnwidth]{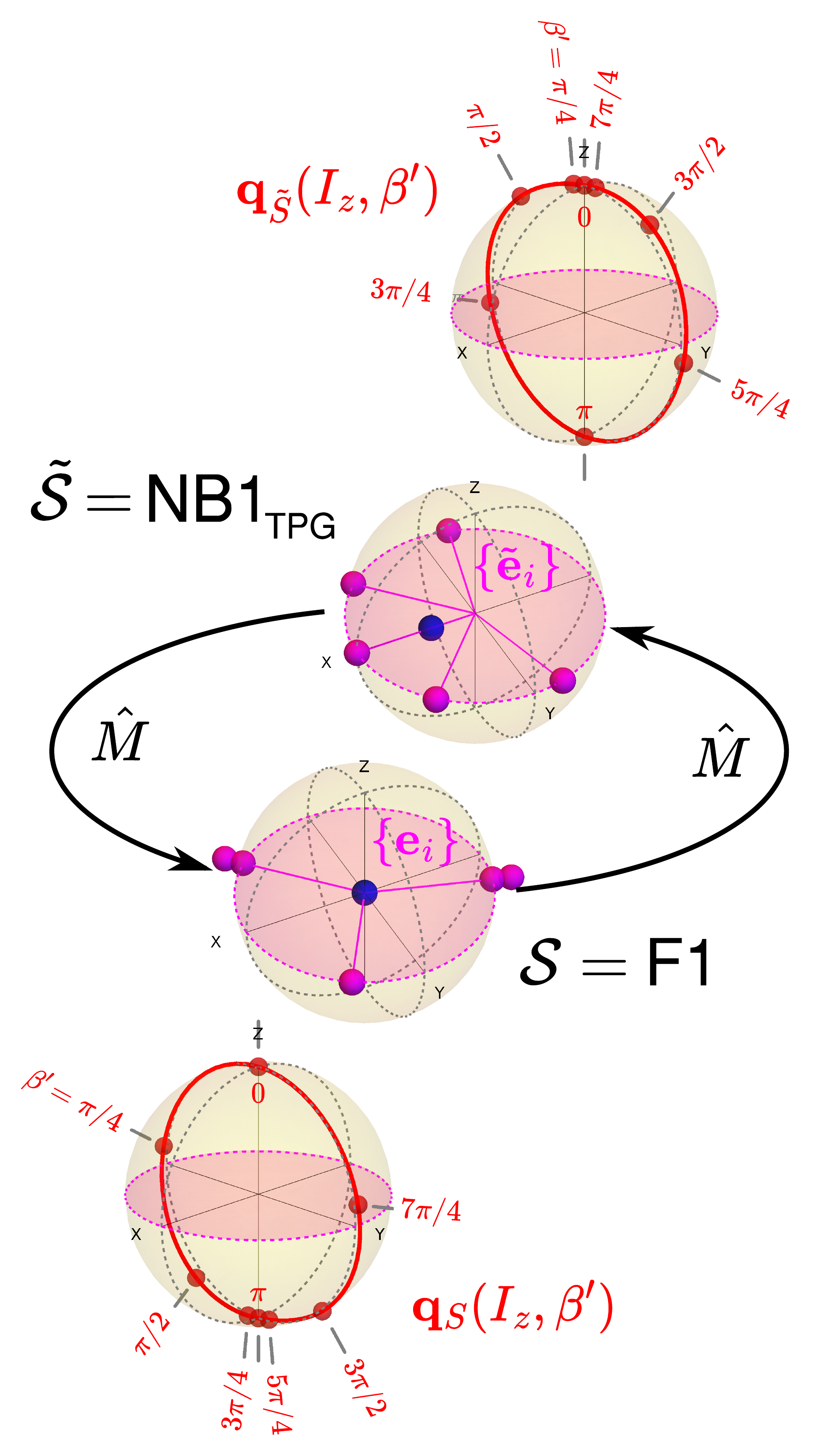}
    \caption{A representation of the universal duality ($m=2$ cycle) between sequences $\mathcal{S}^{(0)}=$ F1 (\autoref{eq:F1definition}) and $\mathcal{S}^{(1)}\equiv$ NB1\textsubscript{TPG} (\autoref{eq:NB1TPGdefinition}).  Magneta lines/spheres show the rotation axes of the piecewise elements of the sequence, while the blue spheres represent the centroids.  At the top/bottom of the figure, the $z$-inversion performance is represented by continuous paths showing the vectors $\mathcal{S}^{(0)}.(0,0,1)$ and $\mathcal{S}^{(1)}.(0,0,1)$ as a function of $0\leq\beta'<2\pi$.  The dual paths are related by a $\pi$ rotation about the $x=(1,0,0)$ axis.}
    \label{fig:F1NB1TPG}
\end{figure}

An archetype of toggling-frame duality is the pair of composite pulse sequences F1 and NB1\textsubscript{TPG}.  F1 \cite{wimperis_iterative_1991,husain_further_2013} is a cyclically permuted version of Wimperis's broadband population-inversion pulse BB1 \cite{wimperis_broadband_1994}, while NB1\textsubscript{TPG} is a reversal-symmetric narrowband inversion pulse devised by Tycko, Pines, and Guckenheimer in the 1980s \cite{tycko_fixed_1985}. 
The sequences are given in the form $\mathcal{S}^{(0)}=(\beta)_{\phi_0^{(0)}}(\beta)_{\phi_1^{(0)}}\ldots(\beta)_{\phi_{n-1}^{(0)}}$ by
\begin{eqnarray}
   \mathrm{F1} &=& (\pi)_{-3\varphi}(\pi)_{-\varphi}\pi_{0}(\pi)_{\varphi}(\pi)_{3\varphi}\,, \label{eq:F1definition}\\
   \mathrm{NB1_{TPG}} &=& (\pi)_{\varphi}(\pi)_{-\varphi}(\pi)_{0}(\pi)_{-\varphi}(\pi)_{\varphi}\,, \label{eq:NB1TPGdefinition}
\end{eqnarray}
with $\varphi=\arccos(-1/4)$, and the common values of $\varphi$ and $n$ already hint at some level of connection between the two.  Applying \autoref{eq:phasetotildeframe} and \autoref{eq:phasefromtildeframe}, we observe 
\begin{eqnarray}
    \phi_{i}^{(1)} \left[\mathrm{NB1_{TPG}}\right] &=& \phi_{i}^{(0)} \left[\mathrm{F1}\right] +4\varphi \label{eq:F1NB1TPGduality} \\ 
    \phi_{i}^{(1)} \left[\mathrm{F1}\right] &=& \phi_{i}^{(0)} \left[\mathrm{NB1_{TPG}}\right] -4\varphi.
\end{eqnarray}
The phase shift of $4\varphi$ arises because in the original definition of NB1$_{\rm{TPG}}$, the net rotation is about the $4\varphi$ axis rather than the $x$ axis.  
The dual properties including the glide-reflection relation of \autoref{eq:glidereflection} are illustrated in \autoref{fig:F1NB1TPG}.

We now identify several other duals, including between (i) antisymmetric narrow-band \cite{vitanov_arbitrarily_2011} and (ii) symmetric broad-band \cite{torosov_smooth_2011,torosov_highfidelity_2011,torosov_arbitrarily_2019} $\pi$-rotation sequences both involving $\beta=\pi$ and phase digitizations $2\pi/n$, introduced by Vitanov and colleagues.  We use modified versions where a global phase shift is added to ensure the phase list is order-reversal antisymmetric.  These modified sequences are denoted with a prime ($\mathcal{N}'_n$ narrowband, $\mathcal{B}'_n$ broadband),
\begin{eqnarray}
\phi_{i} \left[\mathcal{N}'_n\right] &=& \frac{2 \pi}{n}  \left[ (-1)^{i+1} \floor*{
   \frac{i+1}{2}}\ + (-1)^{(n-1)/2} \floor*{\frac{n+1}{4}} \right]\,,\label{eq:defineVitanovNnpulse}\\
 &=& - \phi_{n-1-i} \left[\mathcal{N}'_n\right]\,,
\phi_{i} \left[\mathcal{B'}_{n}\right] \nonumber \\
\phi_{i} \left[\mathcal{B}'_n\right] &=& -\frac{2\pi}{n} \times \frac{2i(i+1)-n^2 +1}{4} \,, \label{eq:defineVitanovBnpulse}
\\
 &=& + \phi_{n-1-i} \left[\mathcal{B}'_n\right]\,, \nonumber
\end{eqnarray}
and produce nominal $\pi$ rotations about the $x$-axis.  Proof that $\mathcal{N}'_{n}$ and $\mathcal{B}'_{n}$ are dual is via \autoref{eqn:DualityOfDifferences}, to show the differences between adjacent phases $\phi_{i} \left[\mathcal{N}'_n\right]$ and $\phi_{i-1} \left[\mathcal{N}'_n\right]$ in the toggling frame equate to those for $\mathcal{B}'_n$.

Other implementations of Vitanov's sequences include scaling the phases by an integer $k$, such that, $\phi_{i} [ \mathcal{N}_{n,k}' ] = k \times \phi_{i} [ \mathcal{N}_{n}' ]$ and $\phi_{i} [ \mathcal{B}_{n,k}' ] = k \times \phi_{i} [ \mathcal{B}_{n}' ]$.  As long as $k$ is not a divisor of $n$, these sequences perform identically to $\mathcal{N}_{n,1}'$ and $\mathcal{B}_{n,1}'$ respectively when considering errors $|\epsilon|>0$, but differently under detuning errors \cite{vandersypen_nmr_2005,torosov_composite_2015}.  Generally, $\mathcal{B}'_{n,k}$ and $\mathcal{N}'_{n,k}$ are also duals. Additionally, Kyoseva and Vitanov \cite{kyoseva_arbitrarily_2013} constructed a recipe for generating band-pass inversion sequences via a nested composition \cite{cho_theory_1987} of broadband and narrowband sequences; these are called $\mathcal{B}_m(\mathcal{N}_n)$ and $\mathcal{N}_m(\mathcal{B}_n)$ and are defined through  
\begin{align}
    \phi[\mathcal{N}_m(\mathcal{B}_n)] = \{
    &\phi_0[\mathcal{N}_m] + \phi[\mathcal{B}_n], \nonumber \\
    &\phi_1[\mathcal{N}_m] + \phi[\overleftarrow{\mathcal{B}}_n], \nonumber \\
    &\phi_2[\mathcal{N}_m] + \phi[\mathcal{B}_n], \nonumber \\
    &\ldots, \nonumber \\
    &\phi_m[\mathcal{N}_m] + \phi[\mathcal{B}_n] \} \,, \label{eq:BinNdefn} \\ 
    & \nonumber \\
    \phi[\mathcal{B}_m(\mathcal{N}_n)] = \{
    &\phi_0[\mathcal{B}_m] + \phi[\mathcal{N}_n], \nonumber \\
    &\phi_1[\mathcal{B}_m] + \phi[\overleftarrow{\mathcal{N}}_n], \nonumber \\
    &\phi_2[\mathcal{B}_m] + \phi[\mathcal{N}_n], \nonumber \\
    &\ldots, \nonumber \\
    &\phi_{m-1}[\mathcal{B}_m] + \phi[\mathcal{N}_n] \} \,,\label{eq:NinBdefn}    
\end{align}
where $\overleftarrow{\mathcal{N}}$ and $\overleftarrow{\mathcal{B}}$ indicate sequences applied in reverse order.  
The linear property $\hat{M} \mathcal{B}(\mathcal{N}) = (\hat{M} \mathcal{B})(\hat{M}\mathcal{N})$ implies that nested dual pulses are also duals, namely that $\mathcal{B'}_m(\mathcal{N'}_n)$ is the dual of $\mathcal{N'}_m(\mathcal{B'}_n)$.

There are many other $\pi$-composite rotation sequences with distinguishing properties, for example the PB1, PB2 \cite{wimperis_broadband_1994} and T1 \cite{wimperis_composite_1989} universal inversion pulses,
\begin{eqnarray}
\text{T1} &=& (\pi)_0 (\pi)_{\varphi_{\text{T1}}}(\pi)_{\varphi_{\text{T1}}}(\pi)_{-\varphi_{\text{T1}}}(\pi)_{-\varphi_{\text{T1}}}\,, \label{eq:T1defn}\\
\text{PB1} &=& (\pi)_{-\varphi_{\text{PB1}}}(\pi)_{-\varphi_{\text{PB1}}}(\pi)_{\varphi_{\text{PB1}}}(\pi)_{\varphi_{\text{PB1}}}\nonumber\\&&\quad (\pi)_{\varphi_{\text{PB1}}}(\pi)_{\varphi_{\text{PB1}}}(\pi)_{-\varphi_{\text{PB1}}}(\pi)_{-\varphi_{\text{PB1}}} (\pi)_0, \label{eq:PB1defn}
\end{eqnarray}
with $\varphi_{\text{T1}}= \arccos{(-1/4)}$, $\varphi_{\text{PB1}}= \arccos{(-1/8)}$. These have the general form $(\pi)_{\phi_a}(\pi)_{\phi_a}(\pi)_{\phi_b}(\pi)_{\phi_b}\ldots(\pi)_{0}$; because axes or phases are repeated twice, $\hat{M}$ automatically maps the vector set onto itself:
$\bm{e}_i^{(1)}\equiv\pm\bm{e}_i^{(0)}$ (or $\phi_i^{(1)}\equiv\pm\phi_i^{(0)}$) and we call these ``half band'' composite rotations, because each $k$\textsuperscript{th} average propagator has the exact same magnitude about both $\beta'=0$ and $\beta'=\pi$.

Returning to the general scenario $\beta=2\pi/m$, where $m$ is an integer, natural axes $\{\bm{e}_i^{(0)}\}$ for composing error-tuned rotation sequences are the vertices of regular shapes with $m$-fold rotational symmetry.  For $m=2$ these are vertices of regular $n$-gons (polygons) \cite{cho_theory_1987,cho_iterative_1987a,vitanov_arbitrarily_2011,genov_correction_2014}, and for higher $m$ the vertices of regular polyhedra -- namely, for $m=3$, a tetrahedron, cube, or dodecahedron, and for $m=4$ and $m=5$, the vertices of a octahedron and icosahedron, respectively.  
Many composite-rotation pulses for example, $90_x 180_y 90_x$, (written equivalently as $90_x90_y90_y90_x$), CORP\textsuperscript{2}SE-$\pi_0$, 
and phase-alternating sequences \cite{shaka_symmetric_1987,ramamoorthy_phasealternated_1991,ramamoorthy_phasealternated_1998,pandey_composite180deg_2015}
involve $\pi/2$ rf-phase steps and octahedral symmetry.  
We say `natural' for the following reasons: (1) All vertices $\bm{e}_i^{(k)}$ belong to the same finite set; (2)
The vertices contain balanced subsets, meaning that $n$ vectors $\{\bm{e}_i^{(1)}\}$ can be chosen to satisfy $\bm{C}^{(1)}=\bm{0}$ and then mapped via \autoref{eq:Rpowerk} to sequences $\mathcal{S}^{(0)}$. 

\begin{figure*}
    \centering
    \includegraphics[width=\textwidth]{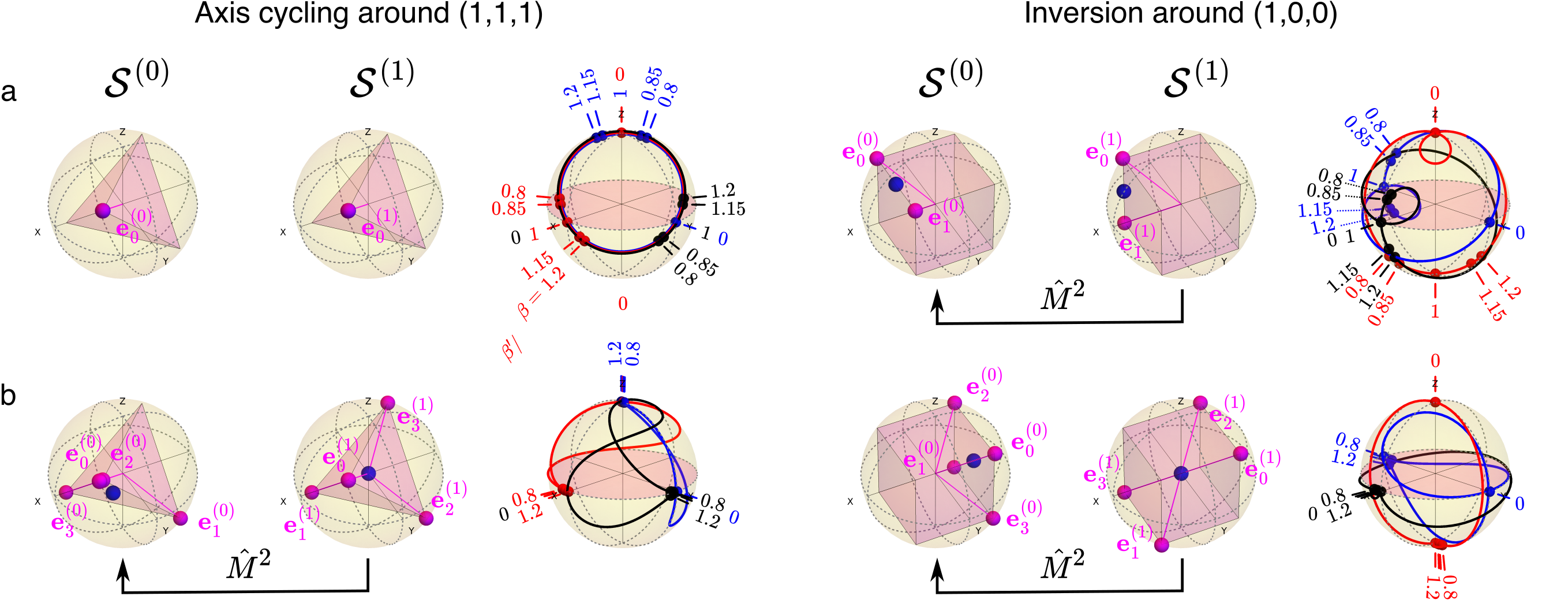}\\
    \includegraphics[width=\textwidth]{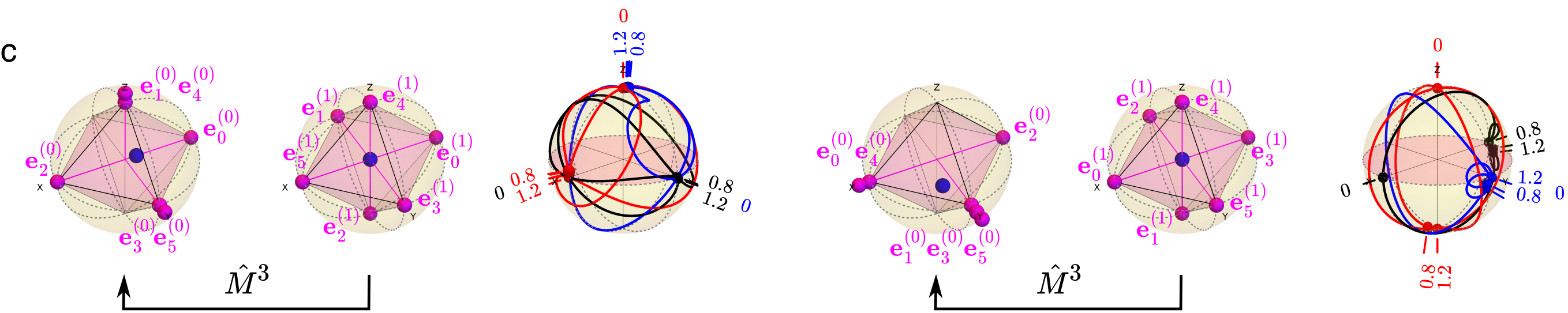}
    \caption{Platonic sequences of composite rotations yielding $(1,1,1)$-axis-cycling or $(\pi)_x$-inversion gates: (a) uncompensated, $\beta=2\pi/3$; (b) compensated, $\beta=2\pi/3$, see \autoref{eq:i34defn} and \autoref{eq:p34defn}; (c) compensated, $\beta=2\pi/4$, see \cite{derome_simple_1988}, \autoref{eq:I46} and \autoref{eq:P46}.  Left/center plots show in 3D the orientations of $\{\bm{e}_i^{(0)}\}$ and $\{\bm{e}_i^{(1)}\}$, using the color convention given in \autoref{fig:F1NB1TPG}.  Right plots show orientation after the sequence as a function of $\beta'$, where the black/blue/red curves denote an initial condition along the x/y/z axis, respectively, and labels indicate positions for selected values of $\beta'/\beta$. 
    }
    \label{fig:period3compositepulses}
\end{figure*}

For $m=3$ and $n=4$, we applied this approach exhaustively and found two original sequences.  First, choosing $\bm{C}^{(1)}=\bm{0}$ for 4 vectors located in a diagonal plane,
\begin{eqnarray}
    \bm{e}_i^{(1)}[{\text I_34(1,0,0)}] &=& \{(-1,1,1), (1,1,-1),\\&&\quad (1,-1,-1), (-1,-1,1)\}/\sqrt{3}\,\nonumber,
\end{eqnarray}
we obtained a compensated $(\pi)_0$ rotation:
\begin{eqnarray}
    \bm{e}_i^{(0)}[{\text I_34(1,0,0)}] &=& \{(-1,1,1), (-1,-1,-1),\\&&\quad (-1,-1,1), (-1,1,-1)\}/\sqrt{3}\,.\nonumber
\end{eqnarray}
which can be compared with two $2\pi/3$ rotations applied sequentially about $(1,1,1)$ and $(1,-1,1)$ (\autoref{fig:period3compositepulses}a+b). 

Another balanced set comprises the four vertices of a tetrahedron, 
\begin{eqnarray}
    \bm{e}_i^{(1)}[{\text P_34(1,1,1)}] &=& \{(1,1,1), (1,-1,-1), \label{eq:i34defn} \\&&\quad (-1,1,-1), (-1,-1,1)\}/\sqrt{3}\,\nonumber,
\end{eqnarray}
and the net transformation is a compensated $(2\pi/3)$ rotation about $(1,1,1)$:
\begin{eqnarray}
    \bm{e}_i^{(0)}[{\text P_34(1,1,1)}] &=& \{(1,1,1), (-1,1,-1),  \label{eq:p34defn} \\&&\quad (1,1,1), (1,-1,-1)\}/\sqrt{3}\,,\nonumber
\end{eqnarray}
or an axis-cycling Clifford gate \cite{crooks_quantum_2024,wolfowicz_pulse_2016}, and is accurate within 5 degrees for up to $\pm20\%$ error in $\beta$ (\autoref{fig:period3compositepulses}a+b).    

For $m=4$ and $n=6$, axes can be chosen as $\{\mathbf{e}_i^{(1)}\} = \{\mathbf{e}_x, \mathbf{e}_{-x}, \mathbf{e}_y, \mathbf{e}_{-y}, \mathbf{e}_z, \mathbf{e}_{-z}\}$ (no specific order) to ensure  balance.  To within a global $z$ rotation, we find the only sequence providing a compensated $\pi_\phi$ rotation is
\begin{eqnarray}
    (\pi/2)_0 (\pi/2)_{\pi/2} (\pi/2)_\pi 
    (\pi/2)_{\pi/2} (\pi/2)_0(\pi/2)_{\pi/2}\,,\label{eq:I46}
\end{eqnarray} 
which is Derome's $(\pi)_y$ sequence \cite{derome_simple_1988,wimperis_shorter_2024}.  Relaxing the balance constraint to $C_z^{(1)}=0$ we also find point-to-point compensated sequences, such as Tycko's amplitude-error-compensated sequence 
$(\pi/2)_0(\pi/2)_0(\pi/2)_{2\pi/3}(\pi/2)_{2\pi/3}(\pi/2)_0(\pi/2)_0 \equiv (\pi)_0(\pi)_{2\pi/3}(\pi)_0$ \cite{tycko_broadband_1983,wimperis_shorter_2024} for $n=6$, and Levitt's offset-error-compensated sequence 
$(\pi/2)_0(\pi/2)_{\pi/2}(\pi/2)_{\pi/2}(\pi/2)_{\pi/2}(\pi/2)_0 \equiv (\pi/2)_0(3\pi/2)_{\pi/2}(\pi/2)_0$ \cite{levitt_nmr_1979,freeman_radiofrequency_1980} for $n=5$.

If sequences with vectors $\mathbf{e}_i^{(0)}$ outside the equatorial plane are permitted, we also find compensated axis-cycling solutions such as
\begin{equation}
\mathbf{e}_i^{(0)}[P_46(1,1,1)] = \{-\mathbf{e}_x,\mathbf{e}_z,\mathbf{e}_x,\mathbf{e}_y,\mathbf{e}_z,\mathbf{e}_y\}\,,   \label{eq:P46}  
\end{equation}
as shown in \autoref{fig:period3compositepulses}a+c, or \begin{equation}
\mathbf{e}_i^{(0)}[P_46'(1,1,1)] = \{-\mathbf{e}_z,\mathbf{e}_x,\mathbf{e}_y,\mathbf{e}_z,\mathbf{e}_x,\mathbf{e}_x\}\,,     
\end{equation}
both for $n=6$. 

In qubit systems, sequences involving non-equatorial axes such as the one given in \autoref{eq:p34defn} can be implemented using near-resonant ac fields with the specific detuning frequencies to set the latitude of the rotation axes $\mathbf{e}_i^{(0)}$.  This approach evidently befits higher-symmetry sets, for example $m=5$ cyclic sequences involving axis subsets of the vertices of the regular icosahedron.  A couple of sequences employing this structure have been used with angles $\beta=2\pi/5$ and $n=10$ or $n=12$ for dipolar-decoupling of diamond nitrogen-vacancy centers \cite{benattar_hamiltonian_2020}. 

Polyhedral symmetry of the axis sets is, however, not always necessary. The design of compensated rotation sequences may capitalize upon relationships between interaction frame cycles for different $\beta$ values, such as between $\beta=\pi/2$ ($m=4$) and $\beta=\pi$ ($m=2$).  
According to \autoref{eq:Rpowerk}, for $\beta = \pi/2$, $\bm{e}_i^{(2)}$ must relate to $\bm{e}_i^{(0)}$ in the same way that $\bm{e}_i^{(1)}$ relates to $\bm{e}_i^{(0)}$ for $\beta = \pi$. 
Known principles follow from this result include the doubling theorems, where: (1) bisection of any $\beta=\pi$ inversion sequence that is symmetric under phase order reversal generates a narrowband $\pi/2$ sequence \cite{wimperis_broadband_1990}; (2) Concatenation of any $\beta=\pi/2$ sequence with its phase-reversed inverse generates a narrowband $\pi$ sequence \cite{husain_further_2013}. 

\begin{figure}
    \centering
    \includegraphics[width=\columnwidth]{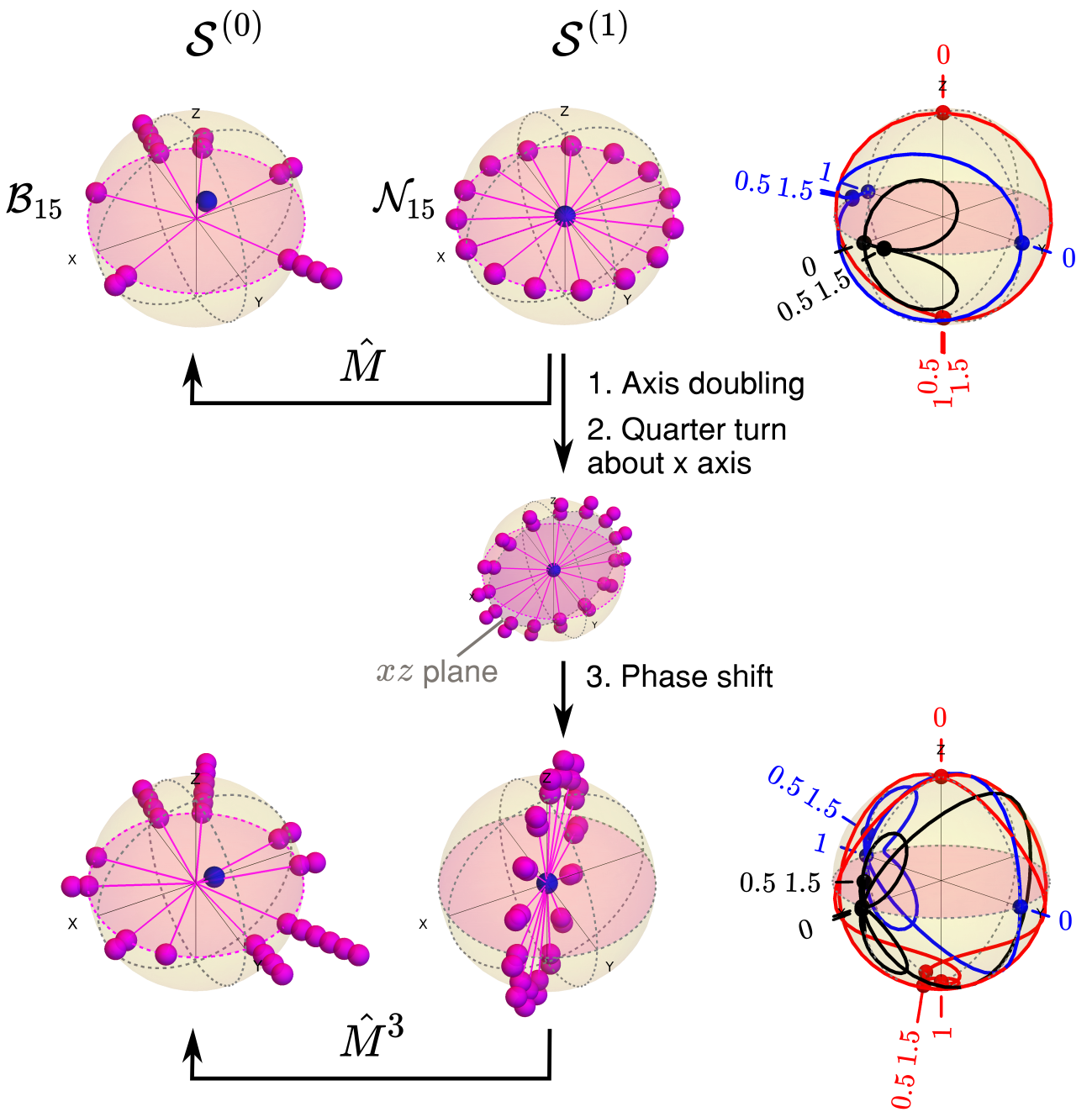}
    \caption{
    Compensated rotation sequences with cycles $m = 2$ and $m = 4$ are interrelated by the general principle where any order-reversal-\textit{symmetric}, broadband inversion sequence composed of $n$ $\beta=\pi$ rotations can be converted to an order-reversal-\textit{antisymmetric} broadband sequence composed of $2n$ $\beta=\pi/2$ rotations.  The example shown is for $n=15$.}
    \label{fig:period4compositepulses2}
\end{figure}

We have used \autoref{eq:Rpowerk} to identify a third connection, hitherto undiscovered, between narrowband universal $(\pi)_x$ rotations composed of units $\beta=\pi/2$ and broadband universal $(\pi)_x$ rotations composed of units $\beta=\pi/2$. Vectors $\{\bm{e}_i^{(1)}\}$ for both sequences are related by a global rotation of $\pi/2$ about the $x$ axis plus a global rotation about $z$.  The global rotation of toggling frame axes does not alter the centroid, since it is already zero. The overall recipe is then: (a) replace every $\pi$ element in a narrowband $(\pi)_x$ sequence with two $\pi/2$ elements of the same phase; (b) rotate each vector $\bm{e}_i^{(1)}$ about $x$ by $\pi/2$, and cyclically permute the sequence order by half its length, (c) apply the $m=4$ inverse-toggling-frame transformation $\hat{M}^3$, and finally (d), rotate each vector $\bm{e}_i^{(0)}$ about the $z$ axis by a constant angle to become reflection symmetric in the $xz$ plane.
This method relates several of the $\beta=\pi/2$ sequences discovered by brute force in Ref.\ \cite{wimperis_shorter_2024} with the analytical solutions of \autoref{eq:defineVitanovNnpulse}.  \autoref{fig:period4compositepulses2} illustrates the relationship for $n=15$, a value for which the $\pi/2$ counterpart would be intractable by brute-force computation.

\textit{Dynamical decoupling (DD)}---Let us now relate these findings to qubit control protocols aimed at noise-suppression and decoherence mitigation, which involve sequences of $n$ $\beta$ kicks that intersperse free-system evolution delays $\tau_j$ ($j\in[0,n]$) \cite{tycko_fixed_1985,merrill_transformed_2014,choi_robust_2020}. 
In the delta-pulse limit, DD sequences are of the form 
\begin{eqnarray}
    S^{(0)} = \tau_{n}\prod_{j=0}^{n-1}R(\beta,\bm{e}_j^{(0)})\tau_j\,,\label{eq:DDseqdefn}
\end{eqnarray}
where axes and delays are tailored for robustness to pulse errors, plus averaging of oscillatory or static external fields. 
\autoref{eq:Rpowerk} has implications in the following scenarios:

(1) \textbf{An oscillating field along the $z$ axis}, of time-dependent amplitude $a\cos(\omega t)$. The propagator of $S^{(0)}$ can be expressed as a product of rotations about axes that are phase-shifted by angles $\theta_j$ depending on $\omega$ and the kick time $t_j=\sum_{l=0}^j\tau_l$:
\begin{eqnarray}
    U_n(S^{(0)}(\omega))&=& R(\theta_n(\omega),\bm{e}_z)\prod_{j=0}^{n-1} R(\beta,\,\hat{R}(\theta_j(\omega),\bm{e}_z).\bm{e}_j^{(0)})\,, \\
    \theta_j(\omega) &=& (a/\omega)\sin(\omega t_j); \quad j\in[0,n-1],\\ 
    \theta_n(\omega) &=& (a/\omega)[\sin(\omega\tau_n)-\sin(\omega \tau_{n-1})] \,.
\end{eqnarray}
Toggling-frame phases are determined using \autoref{eq:phasetotildeframe} and $\phi_j^{(0)}(\omega)=\phi_j^{(0)} +\theta_j(\omega)$.

A non-robust DD sequence that has been used in both liquid-state \cite{jenista_optimized_2009} and solid-state \cite{uhrig_keeping_2007} qubit control is Uhrig decoupling (UDD), a $n$ $\beta=\pi$-pulse scheme that is optimal for highpass noise filtering (giving a filter of cutoff order $n$ in $\omega$, due to design where kick times correspond to the Chebyshev nodes of the second kind on $[0,n]$: $t_j = n\sin^2(\pi j/(2n+2))$, $\bm{e}_j^{(0)}\equiv\bm{e}_x$) but 
is highly sensitive to inaccuracy and inhomogeneity in $\beta$.  The source of the non-robustness is the constant pulse phase $\phi^{(0)}_j=0$, where $\bm{C}^{(1)}\approx\bm{e}_x$ for small $\omega$.  Other DD sequences based on dual-compensated rotations, e.g., URn \cite{genov_arbitrarily_2017}, on the other hand, can be robust to pulse errors but not necessarily noise \cite{kabytayev_robustness_2014}. 

Fully robust DD schemes -- with simultaneous tolerance to $\beta$, rf-detuning and $\omega$ errors -- generally involve concatenation of noise-tolerant and dual-compensated rotation sequences \cite{shaka_iterative_1988, gullion_new_1990}.  One of the most experimentally successful to-date is the $n=20$-pulse ``KDD'' scheme \cite{freeman_decoupling_1997,souza_robust_2011,souza_robust_2012,farfurnik_optimizing_2015} which is a concatenation (or \emph{supercycle} \cite{levitt_symmetry_2008}) of the UR4 sequence XY4 = $\tau/2 - (\pi)_0 - \tau - (\pi)_{\pi/2} - \tau - (\pi)_0 - \tau - (\pi)_{\pi/2} - \tau/2$ \cite{genov_arbitrarily_2017} and the element KDD = $\tau/2 - (\pi)_{\pi/6} - \tau - (\pi)_{0} - \tau - (\pi)_{\pi/2} - \tau - (\pi)_{0} - \tau - (\pi)_{\pi/6} - \tau /2$, which for $\tau=0$ is the Tycko-Pines-Guckenheimer U\textsubscript{5} composite pulse \cite{tycko_iterative_1984,ryan_robust_2010,genov_universal_2020}.
The phases are $\{\phi^{(0)}[{\rm KDD}]\} \equiv \{{\rm \phi^{(0)}[XY4(U_5)]}\}$ following the notation of \autoref{eq:BinNdefn}.  
Other concatenations of URn and dual-compensated rotations, e.g., UR6(U\textsubscript{5})$\equiv$BB1($2\pi$)(U\textsubscript{5}), UR8(U\textsubscript{5}) \cite{cho_frequency_1989}, UR10(U\textsubscript{5}), XY4(U\textsubscript{7}), and so on, extend KDD to arbitrarily improved DD against pulse errors $\beta$.  
We note, historically, that KDD was first reported by the Pines group well before the advent of quantum information processing. Fujiwara and Nagayama, around that time, also presented schemes involving U$_n$ concatenated with MLEV4 = $\tau/2 - (\pi)_0 - \tau - (\pi)_{\pi/2} - \tau - (\pi)_{\pi/2} - \tau - (\pi)_{0} - \tau/2$, which were compensated against rf-pulse error but not noise \cite{fujiwara_composite_1988,fujiwara_frequencyswitched_1993}.  

\begin{figure}
    \centering
    \includegraphics[width=\columnwidth]{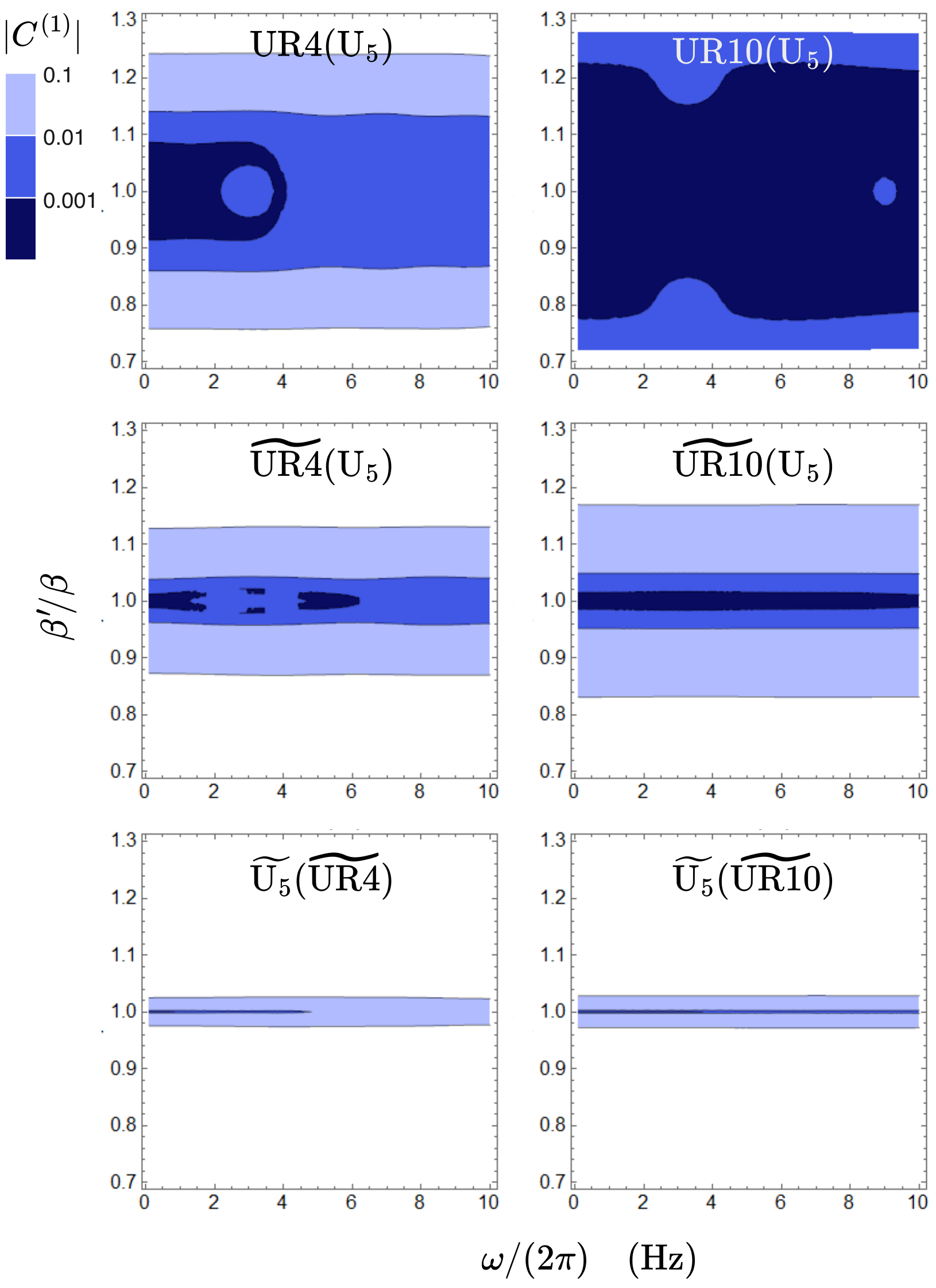}
    \caption{
    Contour-plot representations of the magnitude of the centroid ($|\bm{C}^{(1)}| \propto |(\overline{\Tilde{V}}_n)^{(1)}|$) in the pulse toggling frame for single DD cycles composed of a concatenation between two compensated rotation sequences with $m=2$.  Down each column, we explore the concatenation of duals (indicated by the tilde), which produces sequences tolerant to a wide range of $\omega$, but tunable performance in $\beta'$.  UR4(U\textsubscript{5}) is the well known KDD sequence \cite{freeman_decoupling_1997,souza_robust_2011,souza_robust_2012,farfurnik_optimizing_2015}. 
    }
    \label{fig:DDsequences}
\end{figure}

It is because of the cyclic map for $m=2$ (\autoref{eq:phasetotildeframe}-\autoref{eq:phasefromtildeframe}) that the duals of U$_n$ or ASBO$n$, moreover, can be nested in UR$n$ to produce ``anti-DD'' sequences that are robust in $\omega$ and narrowband in $\beta$, see \autoref{fig:DDsequences}.  We also find that U$_n$ can be substituted by other $n$-odd dual-compensated sequences, such as ASBO$n$$\Omega$ and ASBO$n$B$_1$ \cite{odedra_dualcompensated_2012}, giving XY4(ASBO$n$$\Omega$) and XY4(ASBO$n$B$_1$) as robust sequences.    

The axis-doubling principle used for mapping $m=2\rightarrow m=4$ in \autoref{fig:period4compositepulses2} provides further solutions, such as XY4(U\textsubscript{5}$\vee$U\textsubscript{5}) as a robust DD sequence for $\beta=\pi/2$. Here $\vee$ indicates a self-riffling operation to duplicate the phases: e.g., $\{\phi_a,\phi_b,\ldots\}\vee\{\phi_a,\phi_b,\ldots\} = \{\phi_a,\phi_a,\phi_b,\phi_b,\ldots\}$.

(2) \textbf{A time-independent field of irreducible tensor rank $\lambda$:} 
Consider a field $a_{\lambda,-\mu}T_{\lambda,\mu}$, where the rank $\lambda$ is invariant under global rotation.  The operator part transforms as $R(\beta,\bm{e}_j^{(i)})^{-1}T_{\lambda,\mu}R(\beta,\bm{e}_j^{(i)}) = \sum_{\mu'=-\lambda}^{\lambda} T_{\lambda, \mu'} D^\lambda_{\mu,\mu'}(R(\beta,\bm{e}_j^{(i)}))$, where $\bm{D}^\lambda$ is the $(2\lambda+1)$-by-$(2\lambda+1)$ Wigner matrix, and substitution into \autoref{eq:DDseqdefn} leads to a first-order-average propagator given by
\begin{align}
    (\overline{\Tilde{V}}_n)^{(1)}(a_{\lambda,-\mu}) &= \exp\left(-\mathrm{i} a_{\lambda,-\mu} \sum_{\mu'=-\lambda}^{\lambda} T_{\lambda,\mu'} \kappa_{\lambda,\mu,\mu'}\right), \\
    \kappa_{\lambda,\mu,\mu'} &= \frac{1}{\sum_j \tau_j} \sum_{j=0}^{n-1} \tau_j\, D^\lambda_{\mu,\mu'}(U_j^{-1}). \label{eq:wignerD}
\end{align}
For $\lambda = 1$, this reduces to the familiar vector average in \autoref{eq:AHT1}.

Sequences that cancel $\bm{D}^\lambda$ components by enforcing \( \kappa_{\lambda,\mu,\mu'} = 0 \) are well established, and for $\lambda\geq1$ almost invariably involve polyhedral symmetries.  Solid-state NMR, for example, since the 1980s has used sequences with $\kappa_{2,0,\mu'}=0$ to decouple the secular part of g-tensor anisotropies or pair dipole-dipole couplings -- both rank-2 interactions \cite{waugh_approach_1968}.  An example involving a tetrahedral geometry is virtual magic-angle spinning (MAS) by $\beta=2\pi/3$ about 
$(1,1,1)$ axes \cite{bodneva_suppression_1987,llor_coherent_1995,llor_coherent_1995a}.  Other sequences, such as
WHH-4 = $(\pi/2)_{0} - \tau - (\pi/2)_{\pi/2} - 2\tau - (\pi/2)_{-\pi/2} - \tau - (\pi/2)_{\pi} - 2\tau$,
MREV \cite{mansfield_symmetrized_1971,rhim_enhanced_1973}, 
echo-WAHUHA \cite{choi_robust_2020}, 
BR24 \cite{burum_analysis_1979}, 
etc. \cite{cory_timesuspension_1990} involve $\beta=2\pi/4$ and are octahedral. 
Zero-field analogs developed by Lee, Suter, and Pines achieve the same type of decoupling by suppressing $\kappa_{2,\mu,\mu'}$ without the secular approximation \cite{lee_theory_1987}, and belong to the same respective symmetries.  Other sequences using icosahedral vertices and $\beta = 2\pi/5$ are known to suppress higher-rank terms such as $\kappa_{3,0,\mu'} = 0$ \cite{read_platonic_2025}, or to decouple rank-1 or rank-2 components selectively \cite{benattar_hamiltonian_2020}. Pulse sequence construction using polyhedral symmetry may be viewed as a special case of constructing pulse sequences using the vertices of polytopes \cite{mamone_orientational_2010,jain_quantum_2024} that may have an arbitrary number of dimensions in general.

The selection rules determining which components $\kappa_{\kappa,\mu,\mu'}$ are zero are also well established from group theory \cite{viola_quantum_2002,viola_robust_2003} and are a central feature of symmetry-based sequence design \cite{levitt_symmetry_2008}.  To guarantee $\bm{D}^\lambda(R)=\bm{0}$, the weighted sum in \autoref{eq:wignerD} can be viewed as a finite-sampling implementation of the Peter-Weyl theorem. This may be solved using the orbit-stabilizer theorem, leading to vector sets $\{\bm{e}_i^{(1)}\}$ comprising vertices of a regular polyhedron and its dual polyhedron, weighted by rotation number: e.g., for OEDD ($n=48$) \cite{read_platonic_2025}, $\{\bm{e}_i^{(1)}\}$ comprises cube vertices $\{\pm1,\pm1,\pm1\}$ with weight 3 and rotation angle $\beta_i=2\pi/3$, plus octahedron vertices $\pm\bm{e}_x$, $\pm\bm{e}_y$, $\pm\bm{e}_z$ with weight 4 and $\beta_i=\pi/2$.  Alternatively, for IEDD ($n=120$) \cite{read_platonic_2025}, $\{\bm{e}_i^{(1)}\}$ comprises the twenty dodecahedron vertices with weight 3 and rotation angle $\beta_i=2\pi/3$, plus twelve icosahedron vertices with weight 5 and $\beta_i=2\pi/5$.  For the tetrahedron, which is a self-dual polyhedron, this geometric construction leads to a cube (TEDD, $n=24$, $m=3$ \cite{read_platonic_2025}).  We note that solutions like OEDD and IEDD obey a modified form of \autoref{eq:Rpowerk} where $\beta$ is replaced by $\beta_i$, and the cyclic relationship still holds, with $m$ given by a lowest common multiplier: for OEDD, $m = 12 = \text{lcm}(3,4)$, and for IEDD $m = 15 = \text{lcm}(3,5)$.


\begin{figure}
    \centering
    \includegraphics[width=\columnwidth]{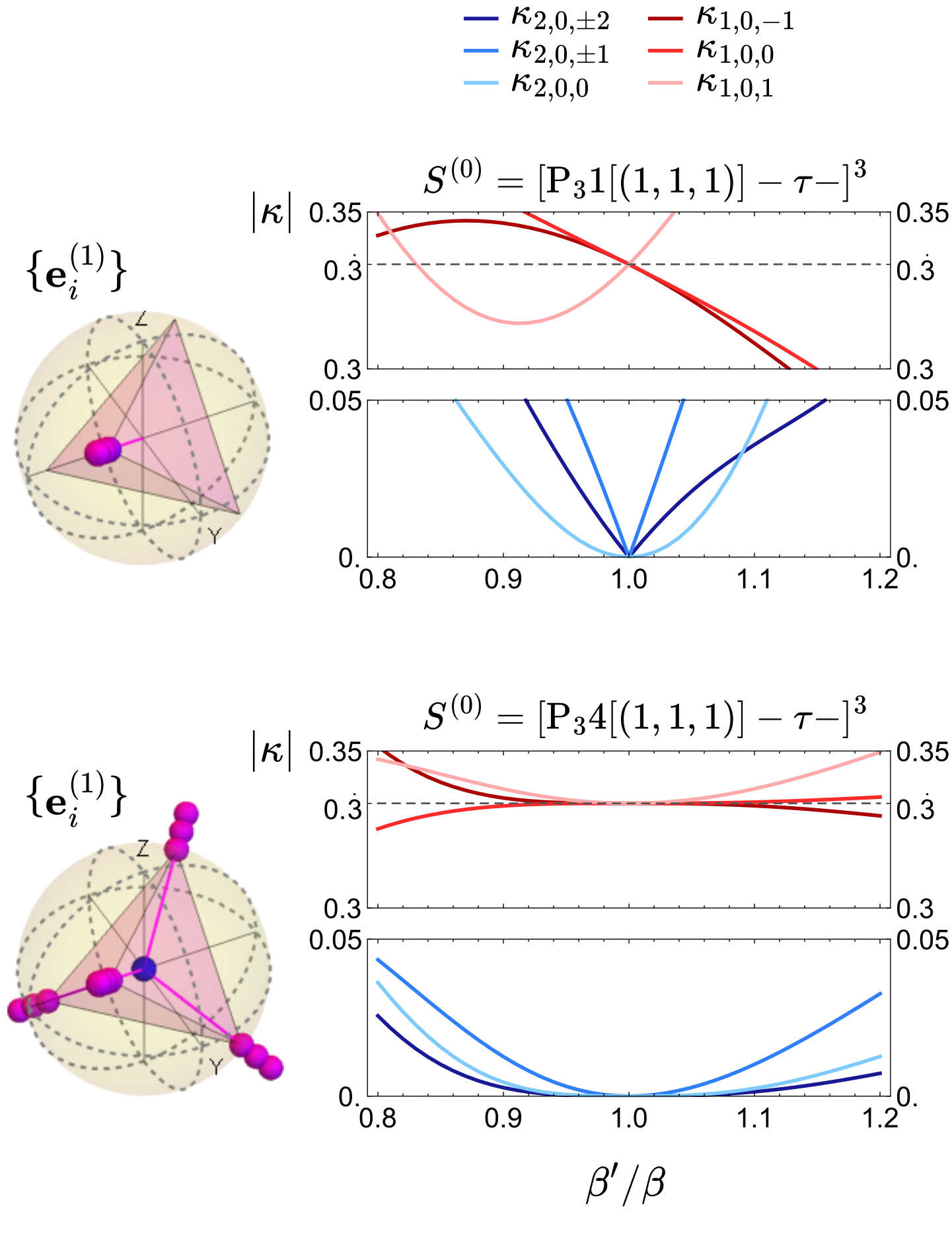}
    \caption{First-order average Hamiltonian suppression of a time-independent background field with rank $\lambda=2$, using:
    (\textit{top}) single $\beta=2\pi/3$ pulses about the $(1,1,1)$ axis as a form of `virtual' magic-angle spinning (MAS);
    (\textit{bottom}) virtual MAS with $(1,1,1)$ pulses compensated for angle error $\epsilon=\beta'-\beta$, defined in \autoref{eq:p34defn}. $\kappa_{\lambda,\mu,\mu'}$ are as defined in \autoref{eq:wignerD}. 
    }
    \label{fig:virtualMAS}
\end{figure}

The sequences above are not inherently robust to errors in $\beta$.  However, replacing each axis-cycling pulse with a compensated rotation can restore the desired averaging even under angle miscalibration.  In \autoref{fig:virtualMAS} we propose a minimal, tetrahedral-symmetry composite pulse for virtual-MAS, where the single axis-cycling rotations are replaced by a compensated rotation of the form given in \autoref{eq:p34defn}.  Whether such constructions are robust to other error sources still remains an open question, however, they may offer a practical path forward.

\textit{Conclusion}---This work identifies cyclic toggling frames for arbitrary sequences of rotations that involve integer fractions of the full circle, namely sequences of rotations about vertices $\{\bm{e}_i^{(0)}\}$ through angles $2\pi/m$: a succession of $m$ iterative transformations always returns $\{\bm{e}_i^{(0)}\}$.  A general formulation of this result appears to be previously unrecognized, and can be used in multiple ways:
First, it leads to the duality of $\pi$-pulse or $m=2$ sequences in pulse-length error.  
The implication of self-mapping vector sets $\{\bm{e}_i^{(k)}\}$ $k \in \{0, \dots, m\}$ is a global property that explains why most error-tolerant composite-rotation and DD sequences -- other than optimal control -- involve angles $2\pi/m$.  For $m=3$ and $4$, this led us to derive several original rotation sequences.  More shall be published in future work.  
We also established direct connections between cycles that relate existing sequence families, particularly between $m=2$ and $m=4$.  

All of these results arguably constitute a unified theory of compensated rotations, because for given values of $n$ and $m$, there is only a finite number of balanced sequences.  For now, however, a more exhaustive analysis of that sort is future work and we close by repeating the appropriate words of Ray Freeman from almost forty years ago  \cite{freeman_handbook_1987}: ``We have certainly not heard the last word on composite pulses.''

\section*{Supplementary material}
A Mathematica notebook file contains the simulation codes used to derive the original sequences, and plot all of the figures presented in this work.
\vspace{1cm}

\section*{End matter}
\appendix*
\textit{Appendix A: Toggling frames for rotation errors.} --- The error-tolerance properties of $\mathcal{S}^{(0)}$ can be analyzed by transforming into alternative reference frames.  Errors in rotation angle (but not axis) may take the general form $\beta\rightarrow\beta'+\epsilon$ for elements of constant nominal rotation angle $\beta'$ and constant or proportional error $\epsilon$, and arise when a spin is addressed by on-resonance ac pulses with a misset Rabi frequency ($\omega_{R,i}$) or length ($\tau_i = \beta/\omega_{R,i}$).  The condition to make $\mathcal{S}^{(0)}$ tolerant to these is to satisfy the approximation $U_n{(\mathcal{S}^{(0)};\beta'+\epsilon)} \approx U_n{(\mathcal{S}^{(0)};\beta')}$ for $|\epsilon|$ below some target threshold.  For this, the toggling frame defined by
\begin{equation}
U_n{(\mathcal{S}^{(0)};\,\beta'+\epsilon)} = U_n{(\mathcal{S}^{(0)};\,\beta')} \Tilde{V}_n(\epsilon)\,,  \label{eq:definitionoftogglingframe2} 
\end{equation}
is convenient because the overall propagator is split into parts that depend separately on $\epsilon$ and $\beta'$.  

The $\epsilon$-dependent part, denoted in \autoref{eq:definitionoftogglingframe2} using a tilde (`$\sim$'), is a sequence of rotations by $\epsilon$, $\tilde{\mathcal{S}}^{(0)} \equiv \{ (\epsilon, \tilde{\bm{e}}_0^{(0)}), \ldots, (\epsilon, \tilde{\bm{e}}_{n-1}^{(0)})\}$, about $n$ axes defined by
\begin{eqnarray}
    \Tilde{V}_n &=& \prod_{i=0}^{n-1} { R(\epsilon, \tilde{\bm{e}}_i^{(0)}) }\,,\\
    \Tilde{\bm{e}}_i^{(0)} &=& (U_i{(\mathcal{S},\beta)})^{-1} \bm{e}_i^{(0)}; \qquad i\geq1\,, \\
    \Tilde{\bm{e}}_0^{(0)} &=& \bm{e}_0^{(0)}\,.
\end{eqnarray}

Frequency-detuning error ($\delta\omega$) is another source of imperfection for ac pulses.  This scenario corresponds to changing the latitude, $\theta_i$, of the vectors $\bm{e}_i^{(0)}$ in the pulse rotating frame to $(\theta_i + \delta\theta_i) = \arctan(\Delta\omega_i + \delta\omega,\omega_{\rm R})$, where $\Delta\omega_i$ is the nominal resonance frequency offset and we assume field oscillates in the $xy$ plane, perpendicular to the main static field axis along $z$. Generally this case presents no simple analytical form for the toggling frame vectors.  However, for $\beta=\pi$ and $\Delta\omega_i=0$, the vectors are simply related to those of the tilde frame, $\{\tilde{\bm{e}}_i\}$, by alternating $\pi/2$ rotations about the $z$ axis \cite{jones_designing_2013,su_quasiclassical_2021}.  We denote this toggling frame with a tilde plus a prime ('): 
\begin{equation}
    \tilde{\bm{e}}_i' \equiv R((-1)^i\pi/2,\bm{e}_z)\,\tilde{\bm{e}}_i\,. \label{eq:Xidefn}
\end{equation}

\textit{Appendix B: Average Liouvillian theory} ---
Provided the error is small, ($|\epsilon/\beta| < 1$), the Baker-Campbell-Hausdorff-Dynkin formula can expand $\Tilde{V}_n$ as a convergent power series in $\epsilon$.  The index $k$ indicates a mean or `average' propagator in $\epsilon^k$, denoted as the $k$\textsuperscript{th}-order average Liouvillian (according to the Nielsen-Levitt convention \cite{hohwy_systematic_1998
} for which $k$ is numbered 1 higher than in earlier literature). 
The lowest $k$ terms have the following form with implicit $\beta'$ dependence in $\{\Tilde{\bm{e}}_i$\}: 
\begin{eqnarray}
    (\overline{\Tilde{V}}_n)^{(1)}(\epsilon) &=& R(\epsilon, \frac{1}{n} \sum_{i=0}^{n-1} \Tilde{\bm{e}}_i)\,, \label{eq:AHT1} \\
    (\overline{\Tilde{V}}_n)^{(2)}(\epsilon) &=& R(\epsilon^2, \frac{1}{2n} \sum_{j=0}^{n-1} \sum_{i=0}^{j} \Tilde{\bm{e}}_j \wedge \Tilde{\bm{e}}_i)\,,\label{eq:AHT2}\\
(\overline{\Tilde{V}}_n)^{(3)}(\epsilon) &=&
R(\epsilon^3, \frac{1}{6n}  \sum_{k=0}^{n-1} \sum_{j=0}^{k-1} \sum_{i=0}^{j-1} \label{eq:AHT3} \\
&&
\left\{ \left[\Tilde{\bm{e}}_i \wedge \Tilde{\bm{e}}_j\right] \wedge \Tilde{\bm{e}}_k + \left[\Tilde{\bm{e}}_k \wedge \Tilde{\bm{e}}_j\right] \wedge \Tilde{\bm{e}}_i \right\}) \nonumber
\\
&&
- \frac{1}{2} \left\{ \left[\Tilde{\bm{e}}_k \wedge \Tilde{\bm{e}}_j\right] \wedge \Tilde{\bm{e}}_j + \left[\Tilde{\bm{e}}_j \wedge \Tilde{\bm{e}}_k\right] \wedge \Tilde{\bm{e}}_k \right\}). 
\nonumber
\end{eqnarray}
and ensemble average is indicated by the overbar ($\overline{\phantom{xx}}$).

Expressions for higher $k$ orders in terms of vector-product sums are generally not needed, but may be obtained by explicit evaluation 
or by recursion \cite{bialynicki-birula_explicit_1969,blanes_magnus_2009}.  

The leading error-compensation properties of $\mathcal{S}^{(0)}$ can thus be linked to \textit{balance} of the vectors sets $\{\bm{e}_i\}$, $\{\Tilde{\bm{e}}_i\}$ and $\{\Tilde{\bm{e}}_i'\}$.  A vector set $\{\bm{v}_i\}$ is balanced if its arithmetic mean or \textit{centroid}  
\begin{eqnarray}
\bm{C}[\{\bm{v}_i\}] &=& (1/n)\sum_{i=0}^{n-1} \bm{v}_i  \\ &\equiv& (C_x[\{\bm{v}_i\}],\,C_y[\{\bm{v}_i\}],\,C_z[\{\bm{v}_i\}]) \nonumber
\end{eqnarray}
is zero, and if the set is not balanced, we say it is \textit{unbalanced}.  Given \autoref{eq:AHT1}, $(\overline{\Tilde{V}}_n)^{(1)} = R(\epsilon, \mathbf{C}[\Tilde{\bm{e}}_i])$, we begin to form a recipe to formulate $\mathcal{S}^{(0)}$: pick a set of balanced vectors, and reverse transform these to the non-toggling frame of rotations.  Generally, a sequence that is broadband to errors $\epsilon$ about $\beta'=\beta$ will have balance $\bm{C}[\{\tilde{\bm{e}}_i^{(0)}\}]=\bm{0}$.  A sequence that is narrowband will be unbalanced about $\beta'=\beta$, $\bm{C}[\{\tilde{\bm{e}}_i^{(0)}\}]\neq\bm{0}$, but balanced about $\beta'=0$ $\bm{C}[\{\bm{e}_i^{(0)}\}] = \bm{0}$.  The second result follows because \autoref{eq:AHT1}---\autoref{eq:AHT3} can be applied using $\Tilde{\bm{e}}_i \equiv \bm{e}_i$, to describe the average propagators around $\beta'=0$, by equivalence to $\beta'=2\pi$.  For detuning errors where \autoref{eq:Xidefn} is valid, balance is required separately for both odd and even-indexed vectors $\tilde{\bm{e}}_i'$ (see for example the U\textsubscript{5} or ``Knill'' pulse described in \cite{tycko_iterative_1984} and  \cite{ryan_robust_2010}).


The second and higher-order error terms in \autoref{eq:AHT2} and \autoref{eq:AHT3}) can be made zero for certain chronological-order symmetries of $\mathcal{S}^{(0)}$.  Known relations comprise: 
(1) order-reversal symmetry \cite{rienstra_efficient_1998,boulant_inhomogeneity_2011} of the sequence, leading to vanishing even order propagators $(\overline{\Tilde{V}}_n)^{(2k)} = 0$ about $\beta'=\beta$ for $\Tilde{\bm{e}}_i = \Tilde{\bm{e}}_{(n-i-1)}$ and similarly $(\overline{\Tilde{V}}_n)^{(2k)} = 0$ about $\beta'=0$ for $\bm{e}_i = \bm{e}_{(n-i-1)}$; 
(2) order-reversal antisymmetry, leading to vanishing of \emph{all} order propagators $(\overline{\Tilde{V}}_n)^{(k)} = 0$ for $\tilde{\bm{e}}_i = -\tilde{\bm{e}}_{(n-i-1)}$.
(3) Nesting, where if $\mathcal{S}^{(0)}$ consists of two or more subunits all individually satisfying $(\overline{\Tilde{V}}_n)^{(1)} = 0$, then the second- and third-order average propagators of $\mathcal{S}^{(0)}$ are simply the product of the second- and third-order average propagators over the individual subcycles, respectively \cite{burum_analysis_1979}.

Combining these properties, widest error tolerance is given when $\bm{C}[\tilde{\bm{e}}_i]=\bm{0}$ and $\mathcal{S}^{(0)}$ is order-reversal symmetric.  

\textit{Appendix C: Superscript notation for successive toggling frames.} --- Here we choose to drop the tilde because $\Tilde{V}_n$ also describes a rotation sequence, albeit through angles $\epsilon$ rather than $\beta$.  Instead, we denote this frame by adding one to the superscript index, so $\Tilde{\bm{e}}_i^{(0)}\equiv\bm{e}_i^{(1)}$.  This is the convention used throughout the main text.  We also note that $\bm{e}_i^{(m)} = \bm{e}_i^{(0)}$ for $\beta'=0$.

\section*{Acknowledgments}
The work described is funded by: 
the Spanish Ministry of Science MCIN with funding from European Union NextGenerationEU (PRTR-C17.I1) and by Generalitat de Catalunya ``Severo Ochoa'' Center of Excellence CEX2019-000910-S; 
the Spanish Ministry of Science projects MARICHAS (PID2021-126059OA-I00), SEE-13-MRI (CPP2022-009771), AMAREI (CNS2023-144106) plus RYC2022-035450-I, funded by MCIN/AEI /10.13039/501100011033; 
Generalitat de Catalunya through the CERCA program;  
Ag\`{e}ncia de Gesti\'{o} d'Ajuts Universitaris i de Recerca Grant Nos. 2017-SGR-1354 and 2021 FI\_B\_01039; 
Fundaci\'{o} Privada Cellex; 
Fundaci\'{o} Mir-Puig;
MS acknowledges support from European Research Council (grant 786707-FunMagResBeacons), 
EPSRC-UK (grant EP/V055593/1).  
We also acknowledge Malcolm Levitt and Morgan Mitchell for discussions. 

\section*{Dedication}
The authors dedicate this paper equally to Malcolm Levitt (on the 45th anniversary of his invention of the first composite pulse), and to Alex Pines (b.\ 1945, d.\ 2024), both pioneers of many topics in NMR -- not least composite rotations.





\bibliography{References/MCDT_MS_CompositePulses}
%

\end{document}